\documentclass{article}

\usepackage{amssymb}
\usepackage{amsmath}
\usepackage{amsthm}
\usepackage{cite}

\newtheorem{Teorema}{Theorem}[section]

\newtheorem{Definizione}{Definition}[section]

\makeatletter
\def\@biblabel#1{#1.}
\makeatother

\begin{document}

\title{Generalized fractional operators\\ for nonstandard Lagrangians\thanks{This is 
a preprint of a paper whose final and definite form will appear in 
\emph{Mathematical Methods in the Applied Sciences}, ISSN 0170-4214. 
Paper submitted 31/Jan/2014; revised 23/Apr/2014; accepted for publication 25/Apr/2014.}}

\author{Giorgio S. Taverna$^1$\\
{\tt tavernaenator@gmail.com}
\and Delfim F. M. Torres$^2$\thanks{Corresponding author.
Tel: +351 234370668; Fax: +351 234370066;  Email: delfim@ua.pt}\\
{\tt delfim@ua.pt}}

\date{$^1$Dipartimento di Fisica, Universit\`{a} di Roma La Sapienza,\\
Piazzale A. Moro 2, 00185 Roma, Italy\\[0.3cm]
$^2$Center for Research and Development in Mathematics and Applications (CIDMA),
Department of Mathematics,\\
University of Aveiro, 3810--193 Aveiro, Portugal}

\maketitle


\begin{abstract}
In this note we study the application of generalized fractional operators
to a particular class of nonstandard Lagrangians. These are typical
of dissipative systems and the corresponding Euler--Lagrange and Hamilton equations are analyzed.
The dependence of the equation of motion on the generalized kernel permits to obtain
a wide range of different configurations of motion. Some examples are discussed and analyzed.
\end{abstract}


\bigskip


\medskip



\section{Introduction}

Fractional calculus plays an important role in the study of different problems in physics,
engineering, finance, and many other branches of science \cite{kil}.
Its formulation dates back to the  $19$th century,
but its applications are, surprisingly, very recent.
One of the most successful results of fractional calculus concerns
the description of anomalous diffusion \cite{mel} and a promising research topic
is the fractional variational calculus \cite{book:frac}.
Close to the fractional variational principle, is the fractional
action-like variational approach (FALVA), where the Lagrangian
(eventually containing fractional derivative terms)
is weighted by a power law function \cite{MyID:85,elna}.
Recent works posed the attention on the application
of a generalized kernel in the action \cite{MyID:226,odz,FVC_Sev,GreenThm,NoetherGen}.
For a survey see \cite{Tat:Del:survey:FVC}.
In this note we follow this approach, applied to nonstandard Lagrangians. Such kind of Lagrangians
cannot be described by a simple difference between kinetic and potential energies,
and are typical to dissipative systems \cite{duffy}. The text is organized as follows:
in Section~\ref{sec:prelim} we review the necessary notions and results from
generalized fractional calculus. Our results appear in Section~\ref{sec:Gen:FEL},
where we obtain and discuss the generalized Euler--Lagrange equations to nonstandard
Lagrangians, considering different kernels. Our results extend those of
\cite{El-Nabulsi:2013,mus}. We end with Section~\ref{sec:conc} of conclusion.


\section{Preliminaries}
\label{sec:prelim}

In this section we briefly review the main notions regarding
generalized fractional operators. For details and for proofs
on the generalized fractional calculus of variations
we refer the reader to \cite{MyID:226,odz,NoetherGen}.
Let us consider a function $l(\tau)$ and another function $k_{\alpha}(t,\tau)$,
called the kernel, eventually depending on $\alpha$.
Throughout the text we assume, if not differently specified, that
\begin{itemize}
\item $0<\alpha<1$;
\item $ t \in [a,b]$;
\item $\tau \in (a,t)$.
\end{itemize}
Following \cite{MR2595959,odz}, we make use of the following definition.

\begin{Definizione}
\label{FractionallAction}
The generalized fractional operator $S_{P}^{\alpha}$ is given by
\begin{equation}
\label{eq:SG}
S_{P}^\alpha[l](t)=p \int_{a}^{t} k_\alpha (t,\tau) l(\tau)d\tau
+q\int_{t}^{b} k_\alpha (\tau,t)l(\tau)d\tau,
\end{equation}
where  $p$ and $q$ are two real numbers, $P=\left \langle a,t,b,p,q \right \rangle$,
and $k_\alpha(t,\tau)$ is the kernel.
\end{Definizione}

It is worth noting that if $P=\left \langle a,t,b,1,0 \right \rangle$ and
$$
k_\alpha(t,\tau)= \frac{1}{\Gamma(\alpha)}(t-\tau)^{\alpha -1},
$$
where $\Gamma$ is the Gamma function, then \eqref{eq:SG} reduces to
\begin{equation*}
S_{P}^\alpha[l](t)= \frac{1}{\Gamma(\alpha)}
\int_{a}^{t}(t-\tau)^{\alpha -1} l(\tau) d\tau.
\end{equation*}
In case $l$ is a Lagrangian, this kind of operator is used to derive
the fractional Euler--Lagrange equations, constituting the so-called
fractional action-like variational approach (FALVA) \cite{MyID:85,elna}.
In this case, $S_{P}^\alpha$ is the left Riemann--Liouville
fractional integral $_{a}I_t^\alpha$ \cite{gor}:
\begin{equation}
\label{eq:RL}
_{a}I_{t}^\alpha[f](t)
=\frac{1}{\Gamma(\alpha)}
\int_{a}^{t} (t-\tau)^{\alpha -1} f(\tau) \,d\tau.
\end{equation}

The following theorem is essential to obtain the Euler--Lagrange equations
in case of generalized kernels.

\begin{Teorema}[Theorem~3.1 of \cite{odz}]
\label{Integration by parts}
Let us consider $k_\alpha$ to be a square-integrable function in
$\Delta= [a,b]\times[a,b]$, $l,m \in L_2 ([a,b])$,
and $P=\left \langle a,t,b,p,q \right \rangle$.
Then $S_{P}^\alpha$ satisfies the following integration by parts formula:
\begin{equation}
\label{eq:PARTS}
\int_{a}^{b} m(t) S_{P}^\alpha [l](t) dt
=\int_{a}^{b}l(t) S_{\hat{P}}^\alpha[m](t) dt ,
\end{equation}
where $\hat{P}=\left \langle a,t,b,q,p \right \rangle$.
\end{Teorema}

If $k_{\alpha}$ satisfies the property $k_{\alpha}(t,\tau)=k_{\alpha}(t-\tau)$,
as is the case for the Riemann--Liouville fractional integral $_{a}I_{t}^\alpha$
\eqref{eq:RL}, then the integration by parts formula \eqref{eq:PARTS}
holds for two functions $l(t)$ and $m(t)$  under hypotheses
as stated in the following theorem.

\begin{Teorema}[Theorem 3.2 of \cite{odz}]
\label{TeoremKernelParts}
If $k_{\alpha}(t,\tau)=k_{\alpha}{ (t-\tau)}$,
$l \in L_1 ([a,b])$ and $m \in C([a,b])$,
then operator $S_{P}^{\alpha}$ satisfies
the integration by parts formula \eqref{eq:PARTS}.
\end{Teorema}


\section{Main Results}
\label{sec:Gen:FEL}

We investigate Euler--Lagrange equations
for actions involving generalized kernels.

\begin{Definizione}
\label{Fractional action}
The generalized fractional action $\mathcal{A}(x)$ is given by
\begin{equation}
\label{eq:FAF}
\mathcal{A}(x)=S_{P_1}^\alpha[L](b)=\int_{a}^{b}
k_\alpha (b,\tau) L(\tau,x(\tau),\dot{x}{(\tau)}) \,d\tau
\end{equation}
with boundary conditions
\begin{equation}
\label{eq:BC}
x(a)=x_a , \quad x(b)=x_b,
 \end{equation}
where $P_1=\left \langle a,b,b,1,0 \right \rangle$
and $L(\tau,x(\tau),\dot{x}{(\tau)})$ is the Lagrangian.
\end{Definizione}

We consider the problem of finding a function $x$ that
minimizes the functional $\mathcal{A}(\cdot)$ subject to boundary conditions \eqref{eq:BC}.
As a corollary of \cite[Theorem~4.2]{odz}, we obtain the following result.

\begin{Teorema}[Generalized fractional Euler--Lagrange equations associated with \eqref{eq:FAF}]
\label{Eulero-Lagrange}
Let $x$ be a solution to the problem of finding a function $x$ that
minimizes the functional $\mathcal{A}$ subject to boundary conditions \eqref{eq:BC}.
If $k_{\alpha}(b,\tau)$ satisfies the conditions of Theorem~\ref{Integration by parts}
or Theorem~\ref{TeoremKernelParts}, together with
\begin{itemize}
\item $L \in C^1([a,b] \times  \mathbb{R}^2 ;\mathbb{R})$,
\item $k_{\alpha}(b,\tau)$, $\partial_3 L \in AC([a,b])$,
\item $k_{\alpha}(b,\tau)$, $\partial_2 L \in C([a,b])$,
\end{itemize}
where $\partial_{i}$ is the partial derivative with respect to the $i$th argument of $L$,
then the following generalized fractional Euler--Lagrange equations hold:
\begin{equation}
\label{eq:EL}
{\partial_2 L}(\tau,x(\tau),\dot{x}(\tau))-\frac {d}{d\tau}\partial_3 L(\tau,x(\tau),\dot{x}(\tau))
=\frac{dk_{\alpha} (b,\tau) }{d\tau}\frac{\partial_3 L(\tau,x(\tau),\dot{x}(\tau))}{k_\alpha (b,\tau)}
\end{equation}
for all $\tau \in [a,b]$.
\end{Teorema}


\subsection{Nonstandard Lagrangians}

When a Lagrangian is expressed as the difference between kinetic and potential energy,
it is called a \emph{standard Lagrangian}. If it is not possible to discriminate
the two contributions of energy, then the Lagrangian is said to be a \emph{nonstandard Lagrangian}
\cite{bas}. An interesting purpose is to find equations of motion able to describe
dissipative dynamical systems by a nonstandard Lagrangian. An equation of motion of form
\begin{equation}
\label{eq:EM}
\ddot{x}(\tau)+A(\tau)\dot{x}(\tau)+B(\tau)x(\tau)=0,
\end{equation}
where $\dot{x}(\tau)=\frac{dx(\tau)}{d\tau}$ and
$A(\tau)$ and $B(\tau)$ are arbitrary, but continuous,
differentiable and integrable functions, is typical of unforced dissipative systems.
As shown by \cite{mus}, equation \eqref{eq:EM} can be derived from
a nonstandard Lagrangian with time-dependent coefficients:
\begin{equation}
\label{eq:NSL}
L(\tau,x(\tau),\dot{x}(\tau))= \frac{1}{r(\tau)\dot{x}(\tau) +s(\tau)x(\tau)},
\end{equation}
where $r(\tau)$ and $s(\tau)$ are continuous and at least twice differentiable functions.
The coefficients $A(\tau)$ and $B(\tau)$ are related to $r(\tau)$ and $s(\tau)$
by the solution of a nonlinear second-order Riccati equation \cite{mus}.
We next obtain, and discuss, the generalized fractional
Euler--Lagrange equations for the nonstandard Lagrangian
with time-dependent coefficients \eqref{eq:NSL}.

Let us take the Lagrangian  \eqref{eq:NSL}
and insert it in \eqref{eq:EL}.
We obtain the equation of motion
\begin{equation}
\label{eq:MG}
\ddot{x} + \frac{\dot{x}}{2}\biggl[\frac{3s}{r}+\frac{\dot{r}}{r}
-\frac{\dot{k}_\alpha}{k_\alpha}\biggr] + {x}\biggl[\frac{s^2}{2r^2}
-\frac{\dot{r} s}{2r^2}+\frac{\dot{s}}{r}
-\frac{s}{2r}\frac{\dot{k}_\alpha}{k_\alpha}\biggr]=0,
\end{equation}
where $\dot{r}(\tau)=\frac{dr(\tau)}{d\tau}$, $\dot{s}(\tau)=\frac{ds(\tau)}{d\tau}$
and $\dot{k}_{\alpha}(b,\tau)=\frac{dk_{\alpha}(b,\tau)}{d\tau}$. This equation
is different from the classical: the fractional version
consists in the presence of the term
$-\frac{\dot{x}}{2} \bigl[\frac{\dot{k}_\alpha}{k_\alpha} \bigr]
-x\bigl[\frac{s}{2r}\frac{\dot{k}_\alpha}{k_\alpha}\bigr]$
\cite{mus}. In contrast, our equation of motion \eqref{eq:MG}
consists of a friction term and a harmonic term, both time depending.
We note that in case $r$ and $s$ are constant in time,
the equation of motion \eqref{eq:MG} reduces to
\begin{equation}
\label{eq:MGC}
\ddot{x} + \frac{\dot{x}}{2}\biggl[\frac{3s}{r}-\frac{\dot{k}_\alpha}{k_\alpha}\biggr]
+ {x}\biggl[\frac{s^2}{2r^2}-\frac{s}{2r}\frac{\dot{k}_\alpha}{k_\alpha}\biggr]=0.
\end{equation}
It is worth noting that if $\frac{\dot{k}_\alpha}{k_\alpha}=\frac{3s}{r}$, then we
get the equation of an undamped oscillator. Nevertheless, this leads to a nonphysical solution,
corresponding to a negative coefficient multiplying $x$. Thus, at least for constant $r$ and $s$,
this kind of Lagrangian describes exclusively dissipative systems. To get the equation of a
damped harmonic oscillator, that is, with coefficients multiplying $x$ and $\dot{x}$ both positive,
it is convenient to consider the sign of the ratio $\frac{\dot{k}_\alpha}{k_\alpha}$.
It is easy to show that if $\frac{\dot{k}_\alpha}{k_\alpha}>0$, then the following condition holds:
\begin{equation*}
0<\frac{\dot{k}_\alpha}{ k_\alpha}<\frac{3s}{r}
\end{equation*}
for all $\tau \in [a,b]$. This poses serious limitations in the behavior of the kernel.
In case $\frac{\dot{k}_\alpha}{k_\alpha}<0$, physical solutions are obtained only if
$\frac{s}{r}>0$ for all $\tau \in [a,b]$. The case $\frac{\dot{k}_\alpha}{k_\alpha}=0$
refers to the classical Euler--Lagrange equations.


\subsection{Hamilton formalism}

Let us consider the Hamilton formalism for a nonstandard Lagrangian $L$ as in \eqref{eq:NSL},
with $r$ and $s$ constant in time and $\frac{s}{r} > 0$. The Hamiltonian is defined by
$$
H(\tau,x(\tau),p(\tau))=p(\tau) \dot{x}(\tau)-L(\tau,x(\tau),\dot{x}(\tau)),
$$
where $p(\tau)= \partial_3 L(\tau,x(\tau),\dot{x}(\tau))$. Being
\begin{equation*}
dH={\partial_1 H}\ d\tau +{\partial_2 H}\ dx+{\partial_3  H}\ dp,
\end{equation*}
where $\partial_{i}$ is the partial derivative with respect
to the $i$th argument of $H$, we have
\begin{equation*}
dH=-{\partial_1 L}\ d\tau-{\partial_2 L}\ dx+\dot{x}\ dp
\end{equation*}
and
\begin{equation}
\label{eq:HAM2}
{\partial_1 H}={\partial_1 L}, \quad
{\partial_2 H}=-{\partial_2 L}, \quad
{\partial_3 H}=\dot{x}.
\end{equation}
At this point, the Hamilton equations follow. The momentum $p$ is
\begin{equation}
\label{eq:P}
p= \partial_3 L = -\frac{r}{(r\dot{x}+sx)^2}
\end{equation}
and the Hamiltonian $H$
\begin{equation}
\label{eq:H}
H=-\frac{sxp}{r}.
\end{equation}
Thus, \eqref{eq:HAM2} can be written as
\begin{equation*}
{\partial_3 H}=-\frac{sx}{r}, \quad
{\partial_2 H}=-\frac{sp}{r}, \quad
{\partial_1 H}=-\frac{s(\dot{x}p+x\dot{p})}{r}.
\end{equation*}
Making use of \eqref{eq:EL} and \eqref{eq:HAM2}, the Hamilton equations for
$\dot{p}=\partial p / \partial\tau$ and $\dot{x}$ are
\begin{equation}
\label{eq:PG}
\begin{split}
\dot{p}&=-{\partial_2 H}-\frac{\dot{k}_{\alpha}}{{k}_{\alpha}}p
=\frac{sp}{r}-\frac{\dot{k}_{\alpha}}{{k}_{\alpha}}p,\\
\dot{x}&={\partial_3 H}=-\frac{sx}{r}.
\end{split}
\end{equation}

It is interesting to note that the Hamiltonian \eqref{eq:H}
is written in a much simpler form than the Lagrangian
\eqref{eq:NSL} and if $r=s>0$, for all $\tau \in [a,b]$,
then the sign of $H$ depends exclusively on $x$ ($H>0$ if $ x>0$, $H<0$ otherwise).
Moreover, it is easy to show that differentiation $p$ \eqref{eq:P} in time
and then developing it in the third member of \eqref{eq:PG}, as in \eqref{eq:P},
it is possible to obtain \eqref{eq:MGC}, that is, the equivalence between
Euler--Lagrange and Hamilton equations is proven.


\section{Conclusion}
\label{sec:conc}

In this brief note we discussed the application of the recent
generalized fractional calculus of variations \cite{MyID:226,odz}
to Euler--Lagrange equations of nonstandard Lagrangians.
The presence of the generalized kernel in the equation of motion modulates
the time dependence of the friction coefficient and of the spring equation,
also in case of Lagrangians with constant parameters $r$ and $s$.
In this case, the Hamiltonian has a sign depending on the momentum $p$
and the coordinate $x$ in a very simple way. This ``comfortable'' Hamiltonian
and the role of the kernel in the Hamilton equations, can be useful
to understand the complex dynamics of dissipative systems. The Euler--Lagrange
equations show a perfect equivalence with the corresponding Hamilton equations,
supporting the procedure suggested in this work. Moreover, the memory effect
in the equation generated by the kernel can avoid insertion of \emph{ad hoc}
time dependent coefficients in the Lagrangian, in order to explain different
kinds of time-dependent dissipative systems.


\section*{Acknowledgements}

The authors are grateful to Roberto Garra for putting them in touch,
and to an anonymous reviewer for valuable comments. The second author
was supported by project PEst-OE/MAT/UI4106/2014 through CIDMA and FCT.



\end{document}